\documentclass[graybox]{svmult}

\usepackage{Setup/polarization}
\makeindex 
\begin{document}
\setcitestyle{authoryear,open={(},close={)}}

\title*{Polarimetry in Planetary Sciences and Astronomy}
\author{C.H. Lucas Patty \orcidID{0000-0002-0073-8879} and\\ Jonathan Grone \orcidID{0000-0001-5074-265X} and\\ Brice-Olivier Demory \orcidID{0000-0002-9355-5165} and\\ Jonas K\"uhn \orcidID{0000-0002-6344-4835} and\\ Jie Ma \orcidID{0000-0003-3583-6652} and\\ Willeke Mulder \orcidID{0000-0001-5762-9385} and\\ Olivier Poch \orcidID{0000-0001-6777-8296} and\\ Antoine Pommerol \orcidID{0000-0002-9165-9243} and\\ Hans~Martin Schmid \orcidID{0000-0002-7501-4015} and\\ Stefano Spadaccia \orcidID{0000-0003-2680-929X} 
}
\date{April 2025}

\institute {C.H. Lucas Patty \at ARTORG Center for Biomedical Engineering Research, Universit\"at Bern, Switzerland,\at Center for Space and Habitability, Universit\"at Bern, Bern, Switzerland, \email{lucas.patty@unibe.ch}
\and Jonathan Grone \at Center for Space and Habitability, Universit\"at Bern, Switzerland 
\and Brice-Olivier Demory \at Center for Space and Habitability, Universit\"at Bern, Bern, Switzerland, \at ARTORG Center for Biomedical Engineering Research, Universit\"at Bern, Switzerland,
\and Jonas K\"uhn \at Division of Space Research and Planetary Sciences, Physics Institute,  Universit\"at Bern, Bern, Switzerland 
\and Jie Ma \at IPAG, Université Grenoble Alpes, Grenoble, France, \at
Department of Physics, ETH  Z\"urich, Z\"urich, Switzerland 
\and Willeke Mulder \at Sterrewacht Leiden, Universiteit Leiden, Leiden, The Netherlands 
\and Olivier Poch \at IPAG, Université Grenoble Alpes, Grenoble, France 
\and Antoine Pommerol \at Division of Space Research and Planetary Sciences, Physics Institute,  Universit\"at Bern, Bern, Switzerland 
\and Hans Martin Schmid \at Department of Physics, ETH  Z\"urich, Z\"urich, Switzerland 
\and Stefano Spadaccia \at Division of Space Research and Planetary Sciences, Physics Institute,  Universit\"at Bern, Bern, Switzerland 
}
\authorrunning{NCCR PlanetS: Polarization}

\maketitle

\section*{Abstract}
In recent decades, the relevance of polarimetry in planetary sciences and astronomy has increased rapidly. Polarization is a fundamental property of light and can be modified by any scattering event. As such, polarization yields additional information that cannot be obtained by only assessing light's scalar properties. For instance, the polarization state of starlight scattered by planetary surfaces can provide useful insights on the composition, size, morphology, and porosity of regolith particles and might even indicate the presence of life. Beside being useful for characterization, polarimetry can also greatly enhance the detection of exoplanets. Here, polarization can be harnessed to enhance the contrast between the bright light of a star, which can be considered to be fully unpolarized, and the very dim but polarized light reflected by an exoplanet. In this paper, we discuss and review the current developments and advances in optical polarimetry and polarimetric instrumentation in Switzerland within the framework of the National Centre of Competence in Research PlanetS. We focus on their implications for the vast range of science cases that polarimetry can address within the research fields of planetary science and astronomy.  
\\
\section{Introduction}
\label{sec:Introduction}
Polarization is a fundamental property of light, and it describes its propagation behavior as a transversal wave. The interaction of light with matter, be it a surface or an atmosphere, will always lead to a degree of polarization. As such, measuring the polarization of light provides additional information beyond standard intensity-based observations.

For example, linear polarization of light can carry information on the surface structure and composition of the material it has been scattered by. In other instances, evaluation of the linear polarimetric phase curve can reveal underlying mechanisms such as shadow hiding \citep{ Hapke1993} or coherent backscatter of the opposition surge \cite{Shkuratov1989, Muinonen1990}. Despite these promising merits, polarimetry is a highly underexploited technique which thus provides a great opportunity ready to be utilized \citep{Snik2013}.

Polarimetry can also be used in life detection. So far, most efforts have been made on the characterization of the scalar properties of light which e.g. can reveal atmospheric biogenic gasses \citep{Kaltenegger2007, DesMarais2002} or pigment signatures \citep{Schwieterman2015}, although its detection is prone to false-positive interpretations \citep{Schwieterman2016,Schwieterman2018}. It has been demonstrated that through spectropolarimetry, life can possibly be detected from afar and without known false positives \cite{Patty2019, Patty2021}. As such, spectropolarimetry can unambiguously indicate the presence of life, and its relation with the scattering geometry provides a unique window to the physical properties of both surfaces and atmospheres that cannot be remotely accessed by any other means.

Over the past decade, polarimetric adaptive optics systems (AO) improved high-resolution imaging observations through fast transformative advances. By correcting for atmospheric turbulence in real time, AO systems allow ground-based telescopes to approach diffraction-limited performance. This opens new frontiers in the study of faint, structured light around bright stars, protoplanetary disks, dusty stellar winds, debris disks. By isolating the polarized component of light from these objects, it is possible to suppress the overwhelming glare of direct starlight and reveal otherwise hidden structures. Additionally, polarimetric AO systems can provide a mean to obtain high spatial resolution data of solar system objects of various sizes.

Within this book chapter, we review and highlight the research conducted within the Swiss National Centre of Competence in Research (NCCR) PlanetS on these highly diverse range of research topics involving polarimetry. In section \ref{sec:Polarization} we give a general introduction in polarization. In section \ref{sec:surface} we describe the polarimetric studies carried out on regoliths and frost. In section \ref{sec:biosig} we describe the studies carried out on using polarimetry to detect biosignatures. The use of polarization in adaptive optics instrumentation is described in section \ref{sec:AO} and we describe the conclusions, outlook and NCCR PlanetS legacy in section \ref{sec:Outlook}.

\section{Polarization}
\label{sec:Polarization}
The polarization state of a wave is specified by the behavior of the electric field vector. If the vector oscillates in only one plane, the light is linearly polarized. If the magnitude of the vector is constant but its direction rotates with time (and the overall behavior can thus be described by a left- or righthanded corkscrew) the light is circularly polarized. Both linear and circular polarization are specific cases of the more general elliptical polarization.

The electric field of a monochromatic electromagnetic wave is described by a complex vector, the Jones vector \(\mathbf{E}\). For a completely polarized monochromatic wave propagating in the \( z \)-direction, \(\mathbf{E}\) is given by:
\begin{equation}
    \mathbf{E} = \begin{bmatrix} E_x \\ E_y \end{bmatrix} e^{i k z},
\end{equation}
where $E_x$ and $E_y$ are the complex amplitudes of the electric field components along the $x$ and $y$ axes, respectively. The relative phase between these components determines the polarization state of the wave. For example, when the components are in phase, the light is linearly polarized, while a phase difference of $\pi/2$ results in circular polarization.

The Jones formalism is restricted to fully polarized light and thus only to non-depolarizing systems. Therefore, it is often used in laboratory experiments as well as in the design and analysis of optical systems where coherence is maintained, i.e. due to its relatively easy applicability. In other fields, where we are dealing with partially polarized light, the Jones formalism is not valid anymore. This is why, e.g. in a remote sensing context, it is necessary to introduce the Stokes-Mueller formalism. Whereas the optical parameters of the Jones formalism are the amplitudes and phases of light, the Stokes formalism deals with light intensities. 

With the electric field vectors $E_x$ along the $x$-axis, perpendicular to the scattering plane (0$^{\circ}$), and $E_y$ along the $y$-axis, parallel to the scattering plane (90$^{\circ}$), and the $z$-axis defined by the propagation axis, the Stokes vector $\mathbf{S}$ is given by:

\begin{equation}
\mathbf{S}=
\begin{pmatrix}
I\\
Q\\
U\\
V\\
\end{pmatrix}=
\begin{pmatrix}
\left\langle E^{}_{x}E^{*}_{x} + E^{}_{y}E^{*}_{y}\right\rangle\\
\left\langle E^{}_{x}E^{*}_{x} - E^{}_{y}E^{*}_{y}\right\rangle\\
\left\langle E^{}_{x}E^{*}_{y} - E^{}_{y}E^{*}_{x}\right\rangle\\
i\left\langle E^{}_{x}E^{*}_{y} - E^{}_{y}E^{*}_{x}\right\rangle\\
\end{pmatrix}=
\begin{pmatrix}
I_{0^{\circ}}+I_{90^{\circ}}\\
I_{0^{\circ}}-I_{90^{\circ}}\\
I_{45^{\circ}}-I_{-45^{\circ}}\\
I_{RHC}-I_{LHC}\\
\end{pmatrix}.
\end{equation}

Again, the Stokes parameters $I$, $Q$, $U$ and $V$ refer to intensities and are thus direct (remotely) measurable quantities. The total intensity is given by Stokes 
\(I\) Stokes \(Q\) and \(U\) represent differences in intensity of linear polarization aligned along specific axes. Specifically, \(Q\) measures the difference in intensity between linear polarizations at \(0^{\circ}\) (horizontal) and \(90^{\circ}\) (vertical), while \(U\) measures the difference in intensity between linear polarization at \(45^{\circ}\) and \(135^{\circ}\). Stokes \(V\) describes the difference in intensity between right-handed and left-handed circular polarization. $I_{0^{\circ}}, I_{90^{\circ}}, I_{45^{\circ}}$ and $I_{-45^{\circ}}$ are the intensities in the respective planes perpendicular to the propagation axis, while $I_{LHC}$ and $I_{RHC}$ are the intensities of left- and right-handed circularly polarized light, respectively, as seen from the observer where $I_{LHC}$ rotates clockwise and $I_{RHC}$ counterclockwise. If the absolute intensity $I$ is known, the polarization state can thus be completely described by the normalized quantities $Q/I$, $U/I$ and $V/I$. The linear ($Q/I$) and circular polarizance ($V/I$) then thus describe the fractionally induced polarization after interaction with any matter.

Although the individual Stokes parameters \(Q\) and \(U\) provide detailed information about the orientation of linear polarization, each is dependent on the chosen reference coordinate system and can vary with viewing geometry. The orientation of linear polarization is typically quantified by the angle of linear polarization (AoLP), defined as:
\[
\phi = \frac{1}{2} \arctan\left(\frac{U}{Q}\right),
\]
where \( \phi \) is the AoLP.

The total degree of linear polarization (DoLP), is a scalar quantity that describes the fraction of the light that is linearly polarized, independent of orientation. It is given by:
\[
\mathrm{DOLP} = \frac{\sqrt{Q^2 + U^2}}{I}.
\]

Within the Stokes-Mueller formalism, any interaction of light with matter can be described by the 4x4 Mueller matrix \(\mathbf{M}\):

\begin{equation}
\mathbf{S}_{\mathrm{measured}}=\mathbf{M}\mathbf{S}_{\mathrm{incident}}=
\begin{bmatrix}
m_{11}&m_{12}&m_{13}&m_{14}\\
m_{21}&m_{22}&m_{23}&m_{24}\\
m_{31}&m_{32}&m_{33}&m_{34}\\
m_{41}&m_{42}&m_{43}&m_{44}\\
\end{bmatrix}\cdot
\begin{pmatrix}
I\\
Q\\
U\\
V\\
\end{pmatrix}_{\mathrm{incident}},
\end{equation}
Consequently, a set of interacting matter in a system can be described by a total system matrix. This matrix is a product of the multiplication of the matrices of \(n\) individual elements in reversed order:
\begin{equation}
\mathbf{M}=\mathbf{M}_{n}\mathbf{M}_{n-1}\ldots\mathbf{M}_{2}\mathbf{M}_{1}.
\end{equation}

Consider a beam of light incident on a detector, with a reference plane perpendicular to the direction of propagation. To analyze the polarization state, we can insert an ideal linear polarizer and rotate its transmission axis within this reference plane. When the polarizer is oriented at \(0^\circ\) and \(90^\circ\), it transmits only the horizontal and vertical linear polarization components, respectively, resulting in measured intensities \(I_{0^\circ}\) and \(I_{90^\circ}\). Similarly, orienting the polarizer at \(+45^\circ\) or \(-45^\circ\) with respect to the horizontal axis yields the intensities \(I_{+45^\circ}\) and \(I_{-45^\circ}\), corresponding to linear polarization at those angles.

To measure circular polarization ($I_\mathrm{RHC}$ and $I_\mathrm{LHC}$) a quarter-wave plate (QWP), a retarder with a retardance $\delta = \pi/2$, is placed before the polarizer with the fast axis aligned along the beam at $\pm 45^\circ$.

A half-wave plate (HWP), retardance of $\delta = \pi$, rotates linear polarization by $2\theta $, where $\theta$ is the angle between the input polarization and the fast axis. This enables selection or modulation of arbitrary linear polarization states when combined with a fixed polarizer.

A polarizing beamsplitter or, e.g., Wollaston prism separates the beam into two spatially distinct, orthogonally polarized components (e.g., $0^\circ$ and $90^\circ$).

\section{Characterization of Regoliths and Frost}
\label{sec:surface}

Linear polarization of reflected sunlight has emerged as an invaluable diagnostic tool to study objects in and beyond our Solar System. The induced linear polarization of the light reflected from a surface or particle can provide valuable information on the composition, shape and size of particles as well as their structures. As such, polarimetric observations provide unique insights that cannot be accessed by measuring the scalar photometric intensity alone.

In order to characterize these surfaces it is important to disentangle the large amount of information contained within the polarization data. A key parameter is the \emph{phase angle} $\alpha$, defined as the angle between the incident and scattered rays (i.e., the `Sun $\rightarrow$ object $\rightarrow$ observer' angle for solar system observations). 

Irregular particles on many Solar System bodies exhibit characteristic polarization-phase curves. At very small phase angles ($\alpha < 3^\circ$), the coherent backscattering opposition effect arises \citep[e.g.][]{Shkuratov1989, Muinonen1990, Hapke1993}, causing the total scattered intensity to increase nonlinearly. Under circularly polarized illumination, this leads to a pronounced surge in reflected circular polarization, whereas linearly polarized illumination experiences a corresponding decrease in reflected linear polarization \citep{Nelson1998}.

The light scattered by a cloud of dust particles/planetary regolith becomes linearly polarized. The linear polarization at small phase angles becomes negative, creating the so-called negative polarization branch (see Fig. \ref{fig:spadaccia}). The phase angle ($\alpha_{\mathrm{min}}$) at which the polarization reaches its minimum,$|P_{\mathrm{min}}|$, usually lies between $8^\circ$ and $15^\circ$. The phase angle at which polarization goes back to 0 and changes sign again is called the inversion angle, $\alpha_\mathrm{inv}$. At larger phase angles ($>30^\circ$), the polarization typically increases until it reaches a maximum $|P_{\mathrm{max}}|$. The maximum in the linear polarization tends to be inversely correlated with the surface albedo, a relationship known as the `Umov effect' \cite{Umov1905}. In many planetary observations, this maximum in linear polarization often appears around phase angles of $90^\circ$--$100^\circ$. 

Indeed, many studies emphasize how especially the negative polarization branch can reveal crucial details about regolith structure and grain properties \citep{Dollfus1989, Bagnulo2015}. Such polarization measurements have historically been used to classify asteroid surfaces and interpret their physical properties, such as grain size distribution and surface porosity \citep{Shkuratov1994, Belskaya2017}.

\begin{figure}
  \begin{center}
\includegraphics[width=0.48\textwidth]{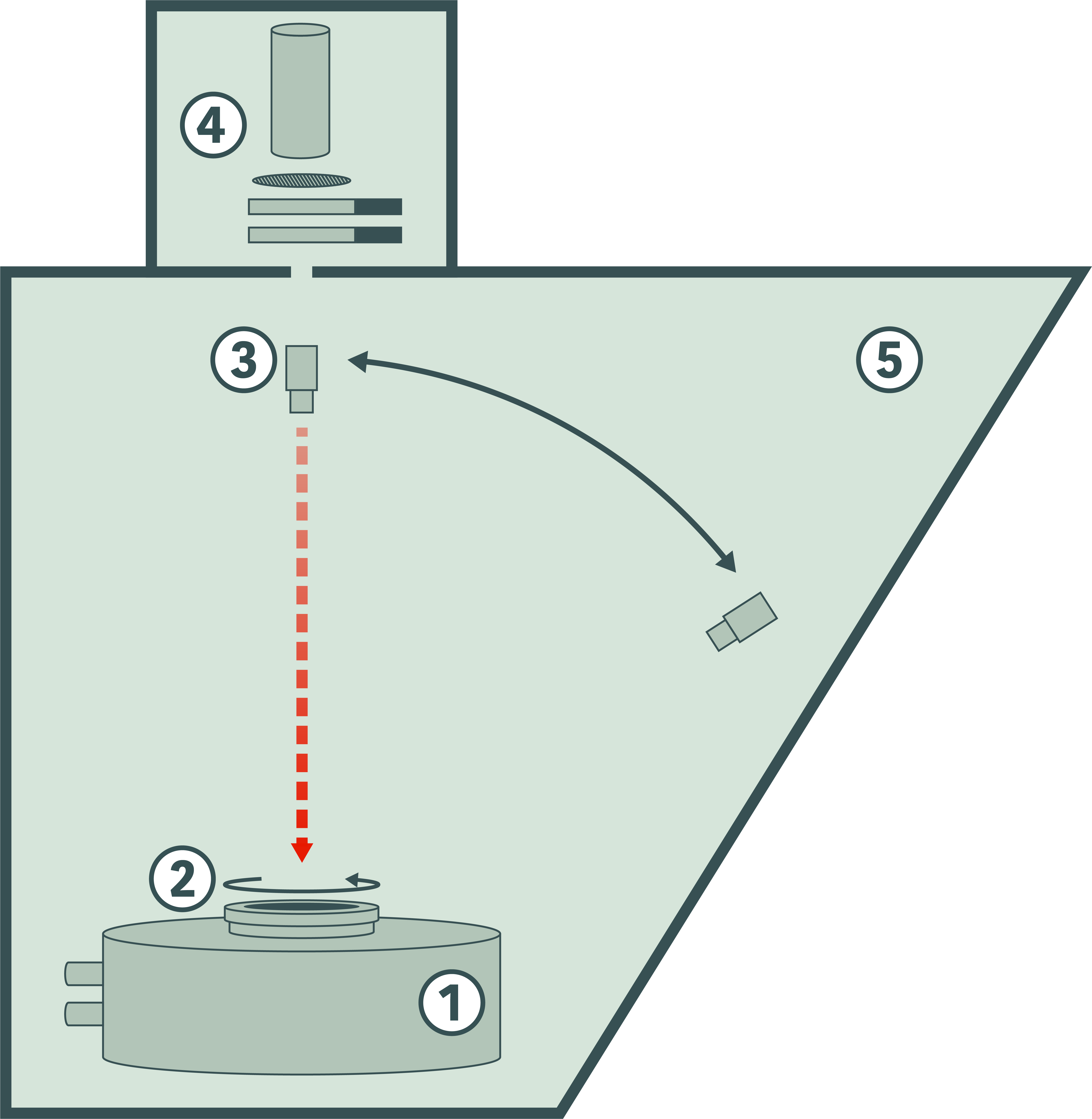}
  \end{center}
\caption{Schematic overview of POLICES. 1) Liquid nitrogen cooled base, 2) sample holder with variable azimuthal angle $\theta$, 3) collimated monochromatic light source with a variable phase angle $\alpha$, 4)  Hinds Instruments Stokes polarimeter in nadir direction, 5) nitrogen flushed measurement chamber.}
    \label{fig:POLICES}
\end{figure}

During the NCCR PlanetS we systematically investigated these effects at the University of Bern, using the POLICES (POLarimeter for ICE Samples) instrument, for a schematic see Figure \ref{fig:POLICES} \citep[see also][]{Poch2018, Patty2022, Spadaccia2023, Spadaccia2022}. The polarimeter (Hinds Instruments II/FS42-47, USA) is based on dual photoelastic modulator (PEM) modulation. Briefly, light from a halogen source and monochromator is depolarized through quartz wedge depolarizers and fiber launched and collimated in a head mounted on a rotating arm. The arm can span a wide range of phase angles from $0.8^\circ$ to $75^\circ$ with a resolution of $0.1^\circ$. Samples are positioned on a rotation stage that can change the azimuth of the sample $\theta$ from $0^\circ$ to $360^\circ$ and are on the same plane that includes the rotation axis of the arm. The entrance pupil of the polarimeter is approximately 50 cm away from the sample, whereafter the light is demodulated through 2 PEMs oriented at 45$^\circ$ modulated at 42 and 47 kHz resonance frequencies, a prism analyzer that is 22.5$^\circ$ to the PEMs and a photomultiplier tube (Hamamatsu R955, Japan).

In one of our recent studies \citep[see][]{Spadaccia2022}, we systematically investigated the polarization properties of common regolith simulants to understand how various mixtures of minerals, with a grain size around 1 $\mu$m, influence the polarization phase curves at small phase angles. Our results show that mixing bright and dark minerals significantly affects the polarization phase curve behavior, particularly causing changes in $|P_{min}|$, $\alpha_{inv}$ and $\alpha_{min}$. 

Our experimental results for instance show that the $|P_{min}|$, as compared to the original endmembers, increases substantially in binary mixtures with different albedoes, like silica-graphite, silica-magnetite, and forsterite-graphite. Similar observations were made by \cite{Shkuratov1987, Shkuratov1994}. This increase caused by the mixture of minerals dominates the negative polarization contribution over other parameters such as particle size distributions, porosity, particles shape, and albedo of the mixed minerals. Our results showed that binary mixtures containing as little as 10 vol\% of a dark component (e.g., graphite or magnetite) caused a significant increase in the absolute value of the minimum polarization $|P_{\min}|$, from ~0.5\% in the pure endmembers up to over 2.2\% in mixtures like silica-magnetite (at a 1:1 ratio, see Figure \ref{fig:spadaccia}). Similar results were also obtained by \cite{Sultana2023} for sub-$\mu$m size grains of olivine and iron sulfide, having albedo and reflectance spectra similar to P/D/X/C- and B-types asteroids. Aggregation (for aggregates size from 2 mm to some cm) does not have any influence on the $P_{min}$ or $\alpha_{inv}$ \citep{Spadaccia2022}. However, the compression of a fine grain sample changes its $P_{min}$ or $\alpha_{inv}$ \citep{shkuratov2002opposition, Spadaccia2022}, indicating an influence of micro-porosity and/or micro-roughness on the polarization.

\begin{figure}
  \begin{center}
\includegraphics[width=0.8\textwidth]{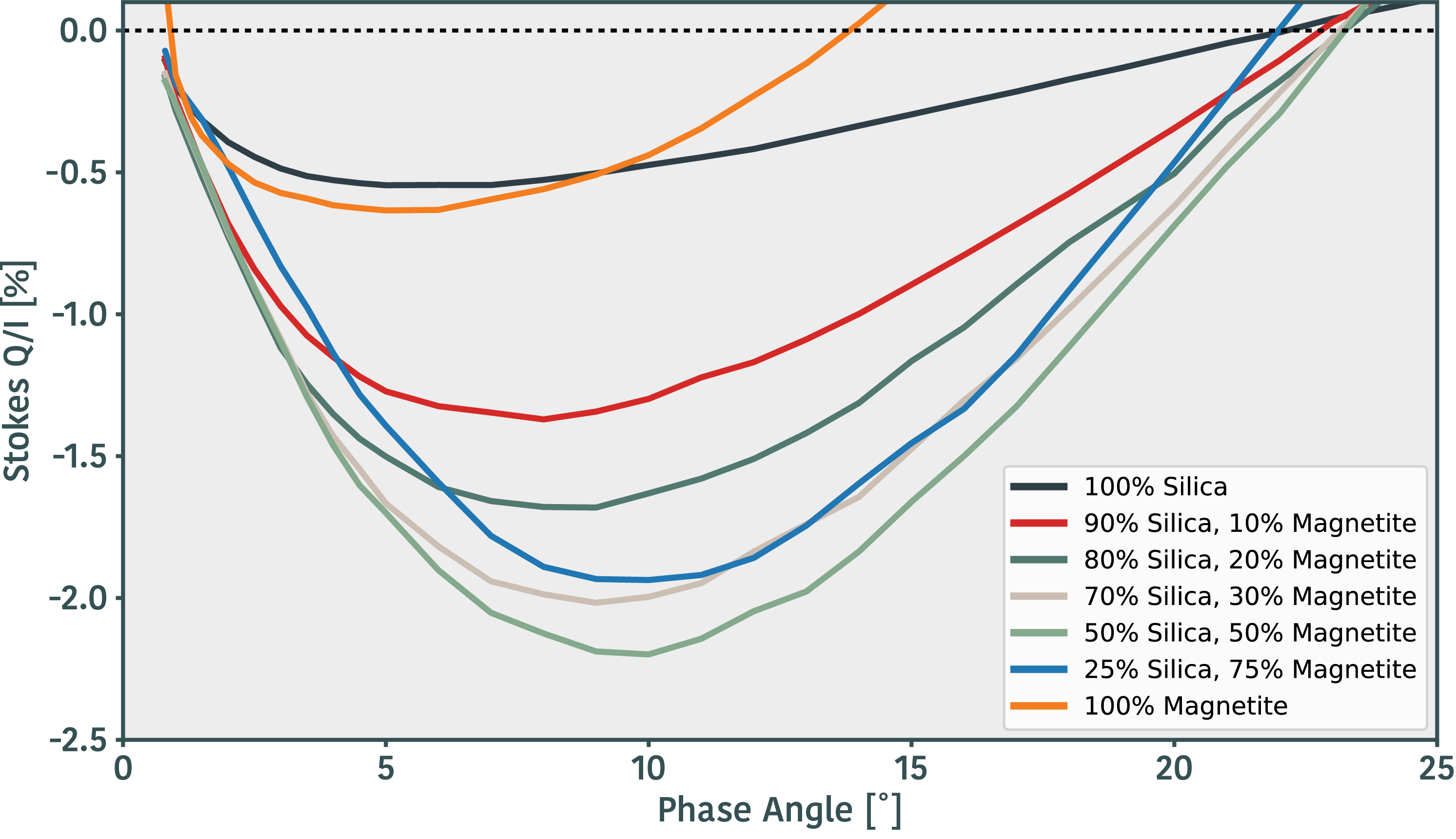}
  \end{center}
\caption{Polarization phase curves of binary mixtures of silica and magnetite. The two mineral components (1$\mu$m) are mixed with mass ratios shown in different colors. The endmembers of the mixture are plotted in black and orange. Credits: Figure adapted from \cite{Spadaccia2022}, licensed under CC-BY 4.0.}
    \label{fig:spadaccia}
\end{figure}

In addition to the binary mixtures, our study on ternary mixtures revealed that polarization phase curves vary notably with the exact mineral ratios, even when maintaining a constant ratio between dark and bright minerals. Such mixtures produce polarization properties that are distinct from all individual components, underlining the complex interplay of particle compositional effects \citep{Spadaccia2022}.

The results in this study are especially significant as they could potentially explain the polarization observations and classifications of certain types of asteroids. The distinctive polarization characteristics of the F- and L-type (Barbarians) asteroids in terms of their $|P_{min}|$ - $\alpha_{inv}$ space can be efficiently explored by mixing bright and dark materials, rather than requiring homogeneous or unique surface compositions. These implications are also not only limited to F- and L-type asteroids but could additionally explain the polarization properties of certain peculiar asteroids that do not belong to a distinctive class, such as Pallas and Lutetia \citep{Spadaccia2022}.

\begin{figure}
    \centering
    \includegraphics[width=0.7\linewidth]{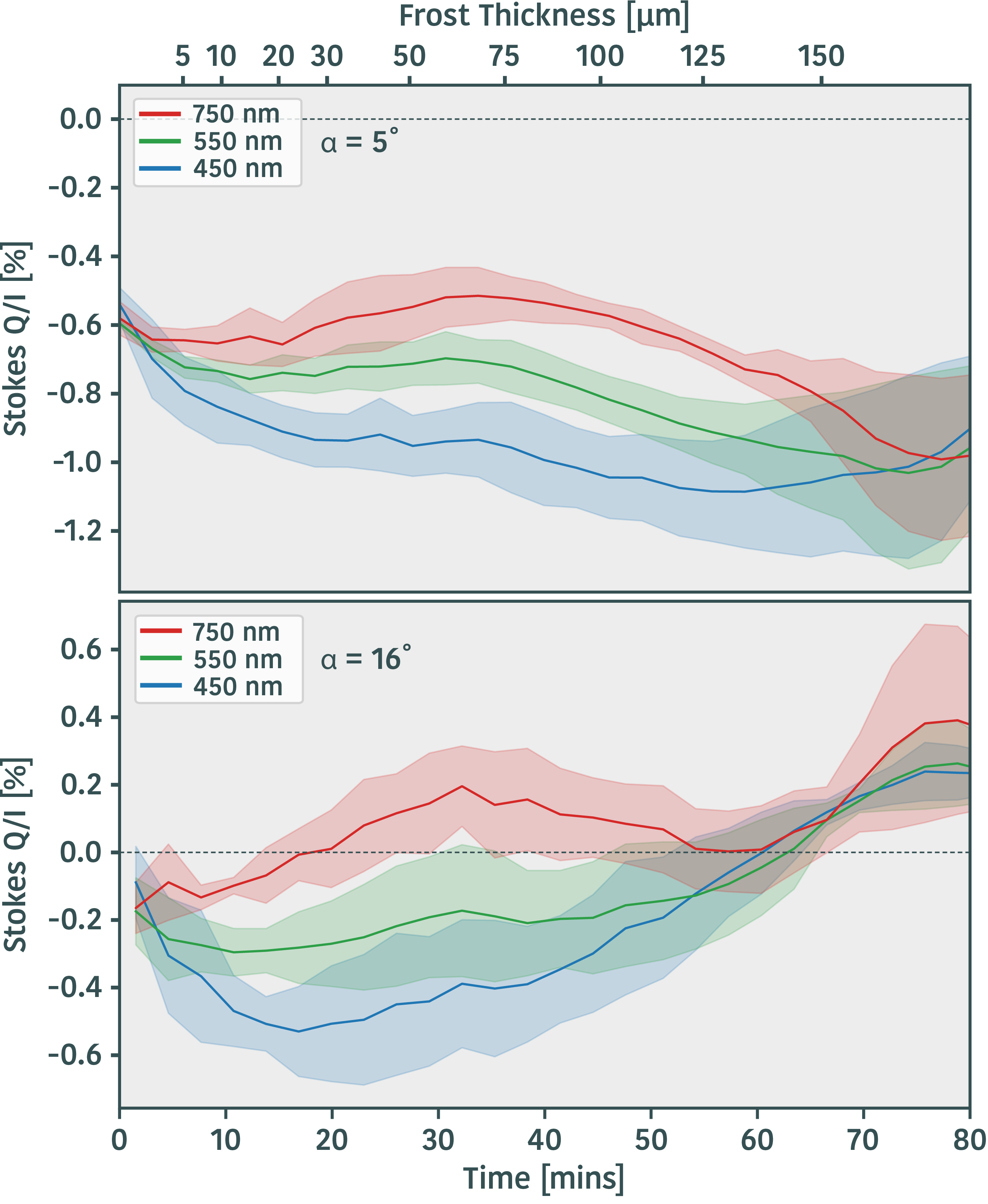}
    \caption{The evolution of the linear polarization at 450, 550 and 750 nm at phase angle of $\alpha = 5^\circ$ (top graph), and $\alpha = 16^\circ$ (bottom graph) and frost
thickness (top bar) on CR regolith simulant. The shaded areas denote the standard deviation of five repetitions. Credits: Figure adapted from \cite{Spadaccia2023}, licensed under CC-BY 4.0.}
    \label{fig:frost}
\end{figure}

Only few studies on the linear polarizance of ice-dust associations exist in literature. At small phase angles, the polarizance of particle surfaces is dominated by multiple scattering between the grains. Once frost forms onto these surfaces, it will alter the scattering environment thus changing the phase curve. \cite{Poch2018} found that fresh frost formed by water condensation at ambient pressure on cold surfaces has a phase curve characterized by resonances (Mie oscillations), indicating that frost embryos are transparent micrometer-sized particles (validated through scanning electron microscopy) with a narrow size distribution and spherical shape, and the angular position of these oscillations was changing with time as the frost gets thicker and more metamorphised \citep[see Fig.\ref{fig:frost} \&][]{Poch2018}. Generally, upon frost formation, the $|P_{min}|$ moves to lower phase angles and at the same time also the $\alpha_{inv}$ decreases in angle. When the frost further accumulates in thickness the negative branch disappears altogether, as depolarization dominates in these highly scattering environments. As such, polarimetric observations at these phase angles could provide a sensitive insight into frost formation. We investigated the detectability of frost fromation on planetary surfaces using both multicolor photometry and multicolor polarimetry \citep{Spadaccia2023}. 

Polarimetric observations at small phase angles near $|P_{min}|$ (5$^\circ$) reveal distinct wavelength-dependent variations, with blue (450 nm) polarization differing significantly from red (750 nm) polarization during the initial stages of frost formation. The maximum observed polarization difference $|P450 - P750|$ ranged between 0.25\% and 0.5\% for frost layers around 10–20 $\mu m$ thick (see Fig. \ref{fig:frost}). This polarization signature decreases substantially when frost crystals exceed 100 $\mu m$, as multiple scattering events within the thicker frost layers randomize polarization states, thus diminishing the wavelength-dependent polarization differences.

Interestingly, linear polarization measurements at a phase angle near $\alpha_{inv}$ (16$^\circ$) exhibit sensitivity to the frost deposition temperature, suggesting a link between the linear polarization signal and the crystalline structure of the deposited frost. Specifically, frost deposited at lower temperatures (-130$^\circ$C) resulted in negative polarization (thus higher $\alpha_{inv}$), while higher temperatures (-100$^\circ$C) favored positive polarization values (thus lower $\alpha_{inv}$). These results are indicative of the different crystalline formations of cubic versus hexagonal ice structures which are temperature dependent. 

As such, using multi-color polarimetry, complementary to photometry, precise information can be acquired remotely about frost deposition and its structure. The results in our study indicate that the combined usage of these techniques potentially can give information on not only the ice deposition and thickness but also on its physical properties and evolution processes.

\section{Detection of Biosignatures}
\label{sec:biosig}
Terrestrial biochemistry is based on a wide range of chiral molecules. In their simplest form, chiral molecules exist in a left-handed (\textsc{l}-) and a right-handed (\textsc{d}-) version, which are non-superimposable mirror images of each other. Unlike abiotic chemistry, where these molecules occur in roughly equal concentrations of both forms (racemic mixture), living organisms almost \citep{Grishin2020} exclusively utilize only one of the mirror-image configurations. The resulting phenomenon whereby molecules of the same type occur in the same chiral form is known as homochirality \citep{Kelvin1904}. For instance, life predominantly synthesizes amino acids in the \textsc{l}-configuration, while sugars implemented in the backbone of nucleic acids are mainly produced in the \textsc{d}-configuration \citep[e.g.][]{Blackmond2011}.

Homochirality also manifests within higher-order structures like biological macromolecules and biomolecular architectures. For example, the $\alpha$-helix, the most prevalent secondary structure of proteins, is almost \citep{Novotny2005} exclusively right-hand-coiled. This homochirality of biomolecules is essential for proper enzymatic function and self-replication, suggesting that it is a prerequisite for life as we know it \citep{Popa2004, Bonner1995, Jafarpour2015}. Thus, it is likely that homochirality is a universal feature of life \citep{Wald1957,MacDermott2012,Patty2018a,Sparks2009} and may serve as an agnostic biosignature of any biochemical form of life.

The molecular dissymmetry of chiral molecules and architectures gives them a distinct response to electromagnetic radiation \citep{Fasman2013,Patty2018a}. This response is most evident when interacting with polarized light, leading to optical rotation and circular dichroism. Circular dichroism, captured by Mueller matrix element $m_{14}$, refers to the differential absorption of left- and right-handed circularly polarized incident light. This absorption produces a net signal only when the molecules are homochiral. Consequently, circular dichroism spectroscopy is widely employed in biomolecular research. Closely related to circular dichroism is circular polarizance, $m_{41}$; the fraction of circular polarization induced when unpolarized light, such as light from a star, interacts with these molecules \citep{Kemp1987}. It has been demonstrated that circular polarization carries the same information as circular dichroism \citep{Patty2017,Patty2018b}, but unlike the latter can be sensed remotely \citep{Pospergelis1969, Wolstencroft2004,Sparks2009,Patty2017}.

Remote spectropolarimetric sensing of Earth's surface could provide novel insights and provide information not accessible through the scalar reflectance alone. The circular polarization response of vegetation, for instance, is a highly sensitive indicator of its physiology, i.e. the macro-organization and the functionality of the photosynthetic apparatus \citep{Garab2009,Garab1996,Lambrev2019}. Consequently, circular polarization measurements could, in principle, detect physiological stress responses, such as drought \citep{Patty2017}, and enable the monitoring of vegetation health \citep{Lambrev2019}. While the circular polarizance of vegetation is generally small (typically less than 1\%), polarization remote sensing could serve as a powerful complementary tool alongside scalar and linear polarization remote sensing for assessing the effects of climate change, desertification, deforestation, and monitoring vegetation health in vulnerable regions.

Beyond vegetation, other phototrophic organisms can also produce circular polarization signals. Some even produce stronger signals than vegetation, reaching up to 2\% in certain multicellular algae \citep{Patty2018c}. Additionally, non-zero polarization has been observed in various oxygenic and anoxygenic phototrophic microorganisms and microbial mats \citep{Sparks2009,Sparks2009a,Sparks2021}.

The discovery of thousands of exoplanets in recent decades (many of them potentially habitable rocky planets) underscores the significance of the fundamental question of finding life beyond Earth. While most research on remotely detectable biosignatures has focused on spectroscopic characterization of planetary atmospheres, the detection of biosignature gases associated with a planetary biosphere is subject tom many false-positive scenarios \citep{Meadows2018a, Schwieterman2016, Schwieterman2018}. Surface biosignatures using polarimetry or spectroscopy, however, are direct observable characteristics of life and have the potential to reduce model dependence \citep{Schwieterman2018}.

Some abiotic materials can create circular polarization in unpolarized light e.g. through multiple scattering by clouds or aerosols. The resulting signals, however, tend to be spectrally flat \citep{Rossi2018}, weaker than those produced by abiotic materials, and lack a directional preference in the absence of structure. Additionally, circular polarization measurements of various minerals, artificial grass fields, and the Martian surface, along with the absence of non-zero signals in these studies, further support the general lack of false positives and abiotic mimicry \citep{Sparks2009a, Sparks2009, Pospergelis1969, Sparks2012, Patty2019}.

\begin{figure}
    \centering
\includegraphics[width=1\textwidth]{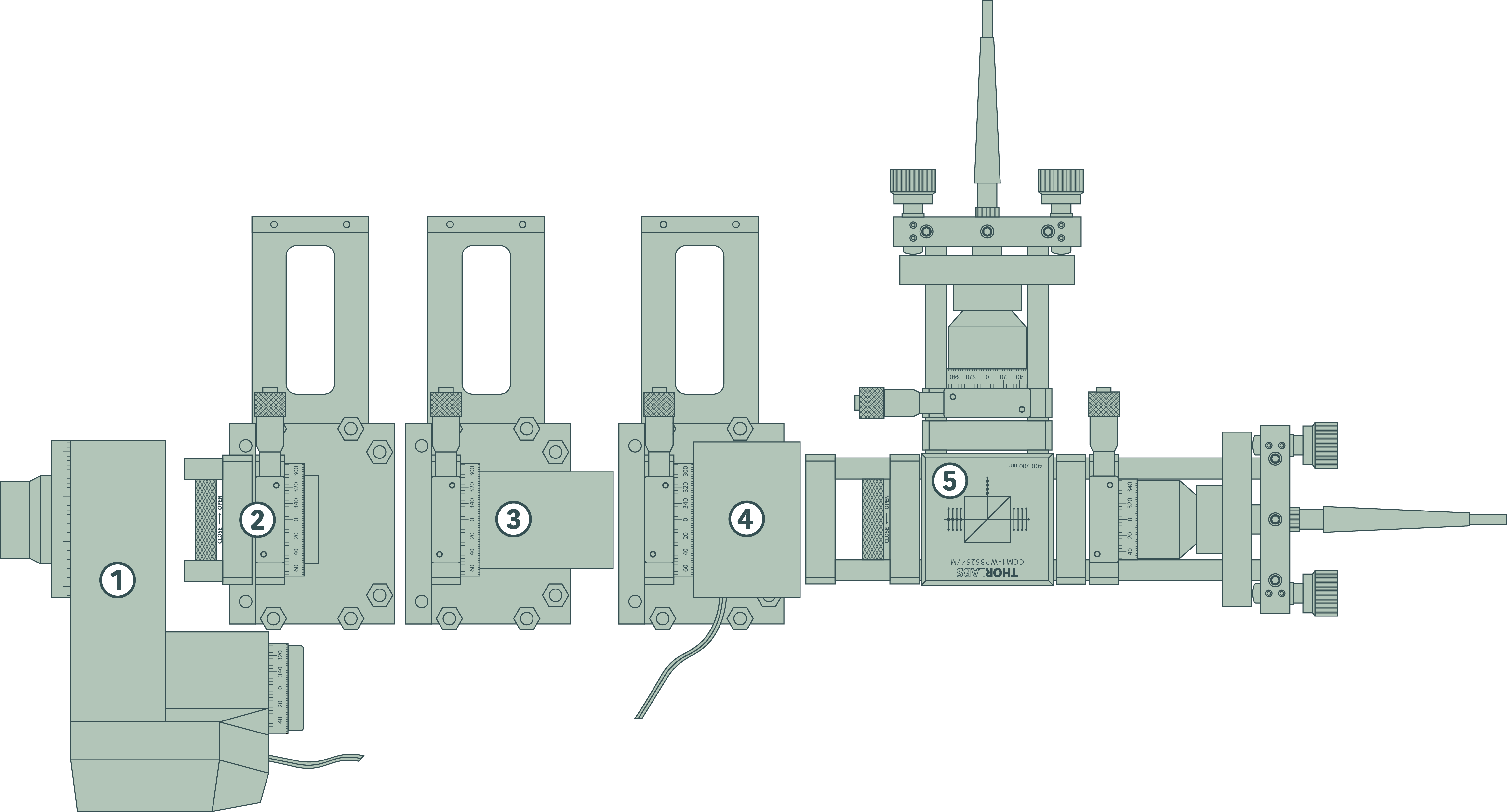}
\caption{Schematic overview of FlyPol. (1) Rotating HWP, (2) QWP, (3) calibration polarizer, (4) FLC, (5) polarizing beam splitter.}
    \label{fig:FlyPol}
\end{figure}

Measuring small, narrowband circular polarization signals produced by living organisms requires dedicated instrumentation. To meet this need, FlyPol was developed within the NCCR PlanetS program \citep{Patty2021} (see Figure~\ref{fig:FlyPol} for a schematic overview). FlyPol is based on the earlier TreePol instrument \citep[see][]{Patty2017,Patty2018a,Patty2019}, but has been specifically adapted for high-sensitivity measurements of fractional circular polarization (Stokes $V/I$).

It measures this signal as a function of wavelength over the 450–800 nm range, achieving high sensitivity ($<10^{-4}$) and accuracy ($<10^{-3}$). This performance is made possible by spectral multiplexing and fast-switching ferroelectric liquid crystal modulation. A dual-beam setup with a polarizing beamsplitter feeds two synchronized spectrographs with orthogonally polarized light, enabling precise and efficient detection.

An achromatic QWP (Fig. \ref{fig:FlyPol} \circled{2}) is used to convert incident circular polarization into linear polarization. This linear component is then modulated by a ferroelectric liquid crystal (FLC) (Fig. \ref{fig:FlyPol} \circled{4}), which functions as a zero-order $\frac{\lambda}{2}$ retarder with a switchable fast-axis orientation of $\pm$22.5$^{\circ}$. Following this, a polarizing beam splitter (Fig. \ref{fig:FlyPol} \circled{5}) separates the orthogonal linear polarization components and directs them into a dual spectrometer setup. To minimize potential cross talk (the unwanted mixing between $Q$, $U$, and $V$ including $I \rightarrow P$), a continuously rotating half-wave plate (HWP) (Fig. \ref{fig:FlyPol} \circled{1}) is placed before the QWP, upstream of all other optical elements. This HWP spins at 5 Hz, effectively averaging out any residual linear polarization effects. 
For FlyPol we have upgraded the spectrographs (Avantes, The Netherlands) to allow for better synchronization with the FLC modulation. Additionally, we incorporated active temperature control for both optical and electronic components. These improvements significantly enhanced the instrument’s polarimetric sensitivity and accuracy, and, crucially, its temporal stability. The latter is of particular importance for obtaining reliable measurements under varying environmental conditions in the field. FlyPol also includes a mechanical actuation system that allows optical elements such as QWP, to be inserted or removed as needed. This enables the sequential acquisition of full-Stokes polarization measurements. The instrument's angular field of view is approximately 1.2$^\circ$. 

\begin{figure}
  \begin{center}
\includegraphics[width=.8\textwidth]{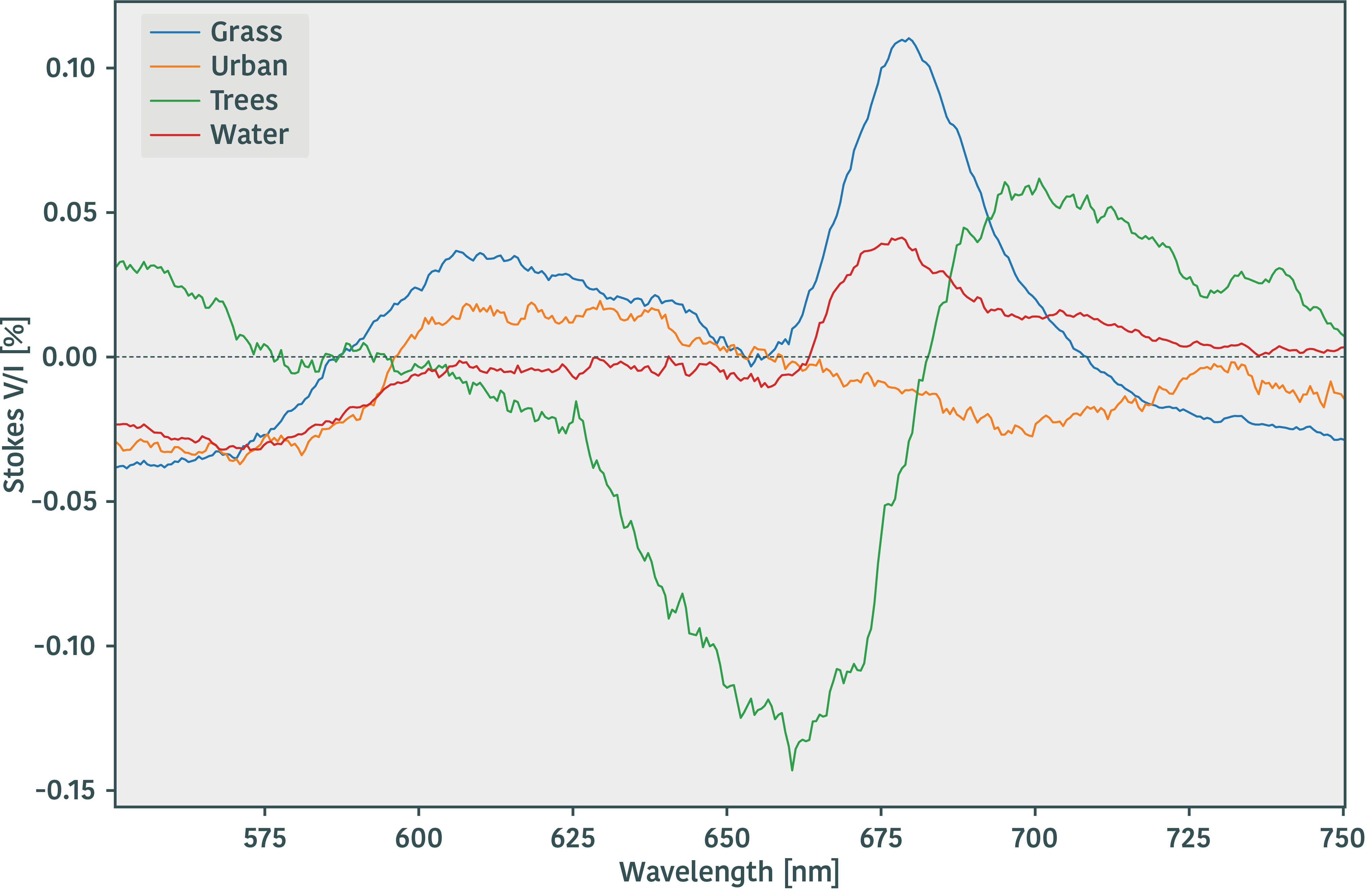}
  \end{center}
\caption{Characterization of different surface types using the FlyPol polarimeter from a helicopter. Credits: Figure adapted from \citet{Patty2021}, licensed under CC-BY 4.0.}
    \label{fig:HelicopterHelicopter}
\end{figure}

Using FlyPol, high-sensitivity circular spectropolarimetric measurements of the Earth’s surface were successfully demonstrated for the first time from a fast-moving aerial platform \citep{Patty2021}. Detection was achieved with integration times as short as one second. The instrument reliably distinguished between urban environments and vegetated landscapes, such as grass fields and forests, with signal-to-noise ratios exceeding 5 and reaching 14, respectively (see Fig. \ref{fig:HelicopterHelicopter}). These results confirm the robustness of circular spectropolarimetry under dynamic remote sensing conditions. As expected, the measured circular polarization signals were subtle.  The difference between the negative and positive band of the vegetation signature (the `circular polarization edge') reached a maximum of $3*10^{-3}$. Additionally, over certain water bodies, circular polarization features were observed that are indicative of the presence of photosynthetic organisms.

A second airborne study was conducted using hot air balloons as reported by \citet{Mulder2022}. Also in this campaign, circular polarization spectra of various landscape elements, including grass fields, bare soil, trees, urban areas, and water bodies were collected. The results qualitatively agree with those obtained during the earlier helicopter-based measurements. However, unlike the previous study, no significant circular polarization signals were detected over water bodies. This discrepancy may be attributed to seasonal differences, which affect biomass, or to differences in observational geometry—such as solar illumination and viewing angles—given that the balloon measurements were performed close to sunset.

\begin{figure}
  \begin{center}
\includegraphics[width=0.98\textwidth]{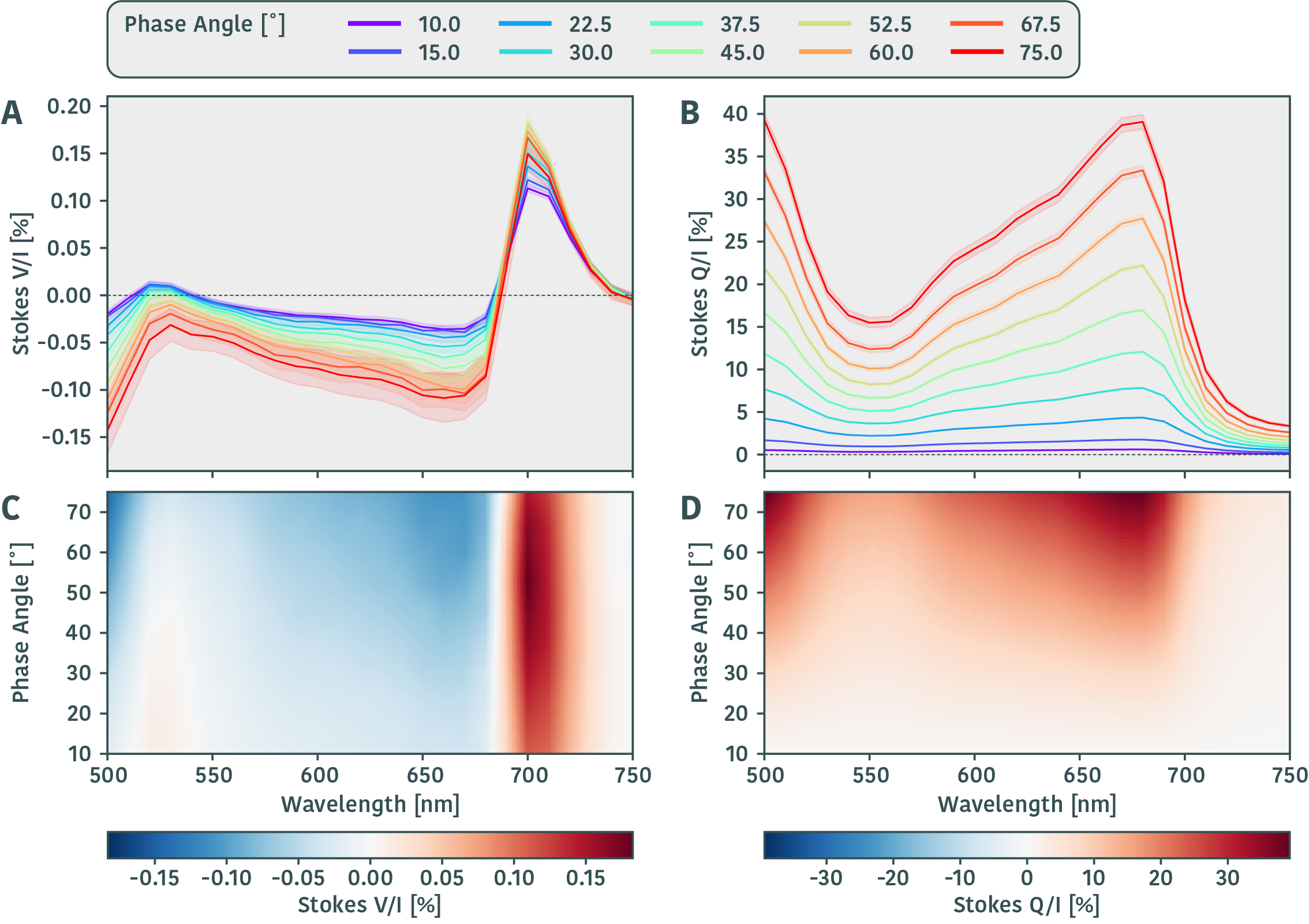}
  \end{center}
\caption{The average polarization spectra V/I (A, C) and Q/I (B,D) of 27 leaves of different species averaged over azimuth. The shaded areas in (A,B) denote the standard error. Credits: Figure adapted from \citet{Patty2022}, licensed under CC-BY 4.0.}
    \label{fig:angles}
\end{figure}

The latter raises an interesting point. While circular polarization signals from vegetation are generally expected to be relatively insensitive to variations in phase angle (see Fig. \ref{fig:angles}A,C) linear polarization effects are strongly affected by scattering geometry (see Fig. \ref{fig:angles}B,D). This is consistent with the origin of the circular polarization signal, which primarily arises from the chiral organization of pigment–protein complexes within plant cells, rather than from scattering processes.

This relationship was investigated in detail by \citet{Patty2022}, who conducted a quantitative and systematic study on the dependency of full-Stokes spectropolarimetric signals on phase angle, using leaves from 27 different plant species (see Fig. \ref{fig:angles}). Measurements were carried out with the POLICES instrument (see Section \ref{sec:surface}, and Fig. \ref{fig:POLICES}). The results confirmed that circular polarizance exhibits limited variability in both magnitude and spectral shape across a range of phase angles. On average, the circular polarization edge reached a peak value of approximately $2.7 \times 10^{-3}$ at a phase angle of $\sim$60$^\circ$, compared to a minimum of about $1.6 \times 10^{-3}$ at 10$^\circ$. However, interspecies variation was more pronounced: some species exhibited circular polarization signals exceeding $V/I > 6 \times {10^{-3}}$ while others showed much weaker signals around $V/I = 1 \times {10^{-3}}$. These findings highlight the relative insensitivity of vegetation circular polarization to phase angle variations, reinforcing its potential as a stable and robust biosignature.

In the same study the linear polarizance of vegetation was also examined. The complementary biosignature arises primarily from scattering at the leaf surface. Similar to the observations of regolith surfaces, the linear polarizance of leaves displayed substantial sensitivity to the phase angle. Specifically,  the difference between the minimum and maximum linear polarizance around the chlorophyll absorption band increasing from approximately 0.5\%at 10$^\circ$ to around 39\% at a phase angle of 75$^\circ$. However, unlike regolith, the linear polarization phase curve of vegetation did not display any negative branch at phase angles above 10$^\circ$. While some interspecies variability in linear polarizance was observed, it was relatively minor compared to the variation observed in the circular polarizance. This suggests that leaf surface scattering is more homogeneous across plant species. 

The combined, simultaneous measurement of both linear and circular polarization signals in remote sensing applications such as aerial platforms, or even orbiters, is highly desirable. Although the circular polarization provides a lot of information on the biota, linear polarization is required to assess surface characteristics and to study habitability indicators such as ocean glint and the occurrence of rainbows \citep{Trees2022}. In addition, full-Stokes polarimetry provides a better handle to investigate systematic effects that are prone to occur in the small circular polarization biosignatures. While the FlyPol instrument demonstrated its capability to measure these biosignatures with high sensitivity, it can not measure all Stokes parameters of the full-Stokes vector simultaneously (only sequentially, which is not favorable for scanning platforms). In addition, the FlyPol design incorporates moving optical components, which are less desirable in space, especially in orbiters.

In a concept by \cite{Sparks2012} both spectroscopy and polarimetry were combined into a snapshot full-Stokes spectropolarimeter. On a 2-D detector, classical spectroscopy would operate in one direction while the polarization would be intensity modulated in the orthogonal direction using a quartz wedge variable retarder. Later work by \cite{Sparks2019} demonstrated the feasibility of this concept, albeit with inclusion of moving parts. A similar solution to the latter was independently found by \cite{Snik2019, keller2020, Mulder2021} in the concept of LSDPol (Life Signature Detector polarimeter), specifically optimized for a high sensitivity in circular polarization using no moving components. LSDpol operates by using a spatial variable liquid-crystal quarter-wave retarder positioned at the instrument’s entrance slit as the primary polarization modulator. This is followed by a fixed QWP, which converts the modulated polarization states between linear and circular. A polarization grating is placed downstream, serving a dual function as both a dispersive element and a polarizing beam splitter. This grating projects two beams with orthogonal polarization states onto a 2-D detector. This enables spectral dispersion along one axis and polarization-dependent intensity modulation along the orthogonal axis. The dual-beam configuration enhances measurement robustness and reduces systematic errors. Despite its promising design, the current prototype unfortunately still experiences notable issues, including polarization crosstalk and spurious polarization modulation caused by Fresnel diffraction effects \cite{Mulder2021}.

\section{Polarization in adaptive optics}
\label{sec:AO}
Adaptive optics (AO) polarimetry has become a powerful tool for high-resolution studies of circumstellar environments. By combining the resolving power of AO systems with the differential sensitivity of polarimetric imaging, astronomers have achieved unprecedented insights into the morphology and composition of protoplanetary disks, debris systems, dust shells, and potentially even extrasolar planets. 

The angular resolution of a telescope is fundamentally limited by diffraction, $R \approx \lambda / D$, where \(D\) is the diameter of the telescope. To achieve the sub-arcsecond resolution at near-infrared and visible wavelengths, large apertures are essential, along with AO systems to correct for atmospheric turbulence that distorts incoming wavefront. AO systems operate in real-time using feedback loop composed of a wavefront sensor, a deformable mirror, and a control computer. Because atmosperic turbulence varies on short timescales ($\sim$1--10 ms), AO corrections must be applied rapidly, which requires a bright wavefront reference source close located near the target. As a result, systems with bright central stars, such as those hosting circumstellar disks or planetary systems, are particulatly well-suited for AO observations.

Polarimetric imaging exploits the differential nature of scattered light: the circumstellar structures polarize and scatter the incident starlight, whereas the light from the star itself is largely unpolarized. This his contrast allows for effective detection of faint structures, even in the presence of residuals from the point spread function (PSF). To enhance this further, modern systems employ coronagraphs to suppress the stellar PSF core, followed by various differential imaging techniques. Under optimal conditions, AO systems like VLT-SPHERE can reach Strehl ratios up to 0.9 in the H-band (1.6~$\mu$m) and 0.6 in the R-band (0.65~$\mu$m) \citep{Fusco2016}. For an 8~m telescope, this corresponds to PSF cores of $\approx$42~mas and $\approx$17~mas in H and R, respectively. Under typical observing conditions and by using faint guide stars (e.g., $>12^m$) these values can be reduced by 30--50\% or more \citep{Milli2017}.

AO polarimeters are classified into three main categories: Double Beam Polarimeters, Integral Field Polarimeters, and Fast Modulation Polarimeters. Each one of them employs specific optical designs and modulation techniques optimized for sensitivity and contrast.

\paragraph{Double Beam Polarimeters}
Instruments such as Subaru-HiCIAO and VLT-SPHERE IRDIS simultaneously record orthogonal polarization states using a polarization beam splitter, typically a Wollaston prism, coupled with a rotating HWP. Simultaneous acquisition of the two polarization states is crucial for mitigating the time-variable speckle noise characteristic of AO observations. By rotating the HWP by 45$^{\circ}$ the sign subsequently switches of the incoming $Q_{in}$- or $U_{in}$-signal, therefore acquiring both $Q^+$ and $Q^-$. The instrumental polarization ($q_{inst}I$) remains unchanged and can thus be compensated by:
\begin{equation}
Q^+ - Q^- = (Q_{in} + q_{inst}I) - (-Q_{in} + q_{inst}I) = 2Q_{in}.
\end{equation}

A switching HWP is a critical component in all AO polarimetric systems. The double-beam modulation technique, adopted by early AO polarimeters such as Subaru-CIAO, VLT-NACO and Subaru-HiCIAO, enables effective suppression of instrumental and atmospheric noise. SPHERE-IRDIS employs a similar strategy, but instead of a Wollaston prism, it uses a non-polarizing beamsplitter in combination with polarizers to mitigate the differential aberrations typically introduced by Wollaston optics \citep{Dohlen2008}. Double-beam polarimeters offer high efficiency and generally have field of view that is larger than for other imaging polarimeter configurations.

\paragraph{Integral Field Polarimeters}
Integral field polarimetry is an innovative method exemplified by the Gemini Planet Imager (GPI). GPI combines integral field spectroscopy (IFS) with polarimetric imaging \citep{Perrin2015}. In this configuration, a two-dimensional array of micro-lenses placed in the focal plane segments the incoming beam into numerous discrete spots. In normal mode, these spots are dispersed with a grating prism (grism), resulting in an array of approximately 36,000 low-resolution spectra on the detector. In polarimetric mode, the grism is substituted by a Wollaston prism. The prism splits each spot into two polarized components with orthogonal polarization states ($I_{\perp}$ and $I_{\parallel}$). This arrangement simultaneously provides spatial and polarimetric information for every lenslet position. Compared to traditional double-beam polarimeters, integral field polarimeters typically have smaller fields of view due to the high spatial sampling density required by the lenslet array.

\paragraph{Fast Modulation Polarimeters}
Lastly, fast modulation polarimetry is a technique that utilizes rapid polarization modulation through ferro-electric liquid crystal retarders, such as in SPHERE-ZIMPOL \citep{Schmid2018}. The polarization modulation is converted into an intensity modulation between orthogonal polarization states (I$\perp$ and I$\parallel$), in synchronous readout by a CCD detector. The main advantage of this technique is its ability to significantly reduce systematic errors caused by speckle noise and flat-field inconsistencies. With a modulation frequency of around 1 kHz, much faster than the typical timescale of atmospheric speckle variations, it enables near-simultaneous detection of I$_\perp$ and I$_\parallel$ on the same detector pixels. This rapid modulation effectively suppresses speckle noise and enhances polarimetric sensitivity  \citep{Schmid2018}. ZIMPOL is optimized explicitly for high-contrast polarimetry, achieving sensitivities down to $10^{-5}$ in the visible wavelength range, making it particularly effective for the search for reflected planetary signals around bright, nearby stars. However, it has comparatively lower polarimetric efficiency (70\% - 90\%), a smaller field of view, and higher requirements for bright guide stars, since the instrument shares light with the visual wavefront sensor.\\

Instrumental polarization effects are generally complicated and need to be accounted for to not compromise the performance of the AO system and the coronograph. Instruments like Subaru-CIAO and Gemini-GPI have a relatively simple design and therefore a relatively simple calibration scheme \citep{Wiktorowicz2014}. This is much harder for instruments located at Nasmyth focus, which is generally favorable for AO, like ZIMPOL and IRDIS.

Inclined optical elements such as mirrors, e.g., the M3 mirror in Nasmyth-mounted systems, can introduce significant polarization effects due to differential reflection and therefore require polarimetric corrections \citep{Joos2008}. Also, image derotators cause polarization cross-talk, converting linear polarization into circular and vice versa, and as such can strongly reduce measurable linear polarization depending on the derotator mirror orientation \citep{Deboer2020}. 

ZIMPOL compensates for these systematic effects with three additional HWPs and a polarization compensator. The latter is neutralizing telescope-induced signals and derotator effects in real time \citep{Bazzon2012}. IRDIS relies primarily on a data reduction package for post-observation corrections using a detailed Mueller matrix model of the instrument calibrated against unpolarized stars \citep{vanHolstein2020}.

To circumvent noise bias, the azimuthal polarization parameter $Q_\phi$ is frequently employed. $Q_\phi$ measures polarization projected onto the azimuthal direction around the central star, such that the $U_\phi$ component is minimized, and the $Q_\phi$ component directly approximates polarized flux avoiding the noise bias: $Q_\phi \approx p_{cs} \times I_{cs}$ \citep{Schmid2006a}. 

Limited observational resolution also impacts quantitative polarimetric analysis due to spatial smearing and orthogonal polarization cancellation effects \citep{Schmid2006a}. The effects largely depend on the atmospheric conditions, but if the PSF is well characterized, these effects can be modeled and accounted for \citep{Tschudi2021}.

\begin{figure}
    \centering
    \includegraphics[width=0.7\linewidth]{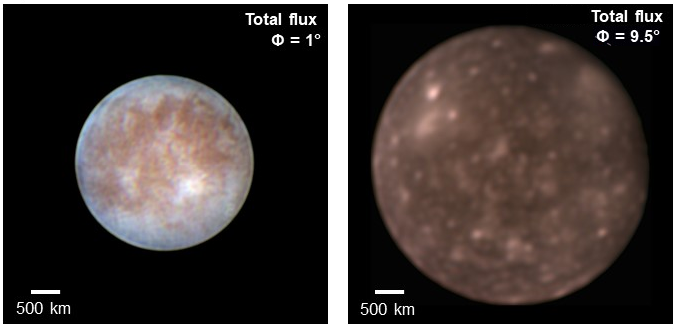}
    \caption{Photometric observations of the Galilean moons Europa (left) and Callisto (right) using SPHERE-ZIMPOL. Figure adapted from Poch, Schmid et al, \textit{in prep.}}
    \label{fig:Zimpol}
\end{figure}

Polarimetric AO imaging has provided key astrophysical insights into various targets within our Solar System (see Figure \ref{fig:Zimpol}) or beyond. Among the most successful targets of AO polarimetry are protoplanetary transition disks. Polarization fractions (Q$_\phi$/I$_\star$) reach ~1\%, showing a high diversity in morphologies such as spiral arms, inner disk cavities, and shadows from inclined disks that are interpreted as planet formation signatures. Fractional polarization ranges from 20–30\% (visual) to 30–60\% (near-IR) at scattering angles around 90° \citep{Monnier2019, Hunziker2021, Tschudi2021}. 

Within the framework of the NCCR PlanetS, \citet{Ma2023} analyzed SPHERE-ZIMPOL and IRDIS polarimetric observations of the transitional disk around RXJ 1604. After correcting for PSF smearing, they recovered intrinsic polarization signals with high fidelity, reporting a maximum degree of polarization of $\sim$40\% in the near-IR. By modeling the photo-polarimetric data, they constrained key dust scattering parameters, such as the asymmetry parameter $g$ and the scattering polarization $p_{\mathrm{max}}$. This study demonstrates the potential for accurate photopolarimetric measurements to determine dust scattering parameters.

In another study, \citet{Ma2024} conducted a systematic analysis of 11 well-resolved protoplanetary disks using SPHERE-ZIMPOL and IRDIS, covering wavelengths from 0.62 to 2.2\,$\mu$m. They applied a novel PSF smearing correction method, enabling the retrieval of $Q_\phi/I_{star}$ with an accuracy of $\sim$10\%. Their results revealed a difference in the polarized color behavior. Herbig Ae/Be disks typically exhibit red polarized colors, pointing to scattering by compact grains. Most T-Tauri disks, on the other hand, tend to appear gray, thereby disfavoring models using highly porous aggregates. Notably, a subset of T Tauri disks displays particularly red polarized spectra, suggesting additional reddening by small inner-disk dust.

Debris disks, composed of dust from planetesimal collisions, are optically thin and typically show lower polarization contrasts ($<0.05\%$). Edge-on debris disks, for example HR 4796A, are ideal for the determination of the dust scattering phase function for both intensity and polarization because the photons undergo only one scattering and the scattering angles are well defined \citep{Milli2019, Chen2020}. As such, they are important for complex dust scattering model validations. 

Dust formation and radiation pressure on dust grains are key processes driving mass loss in red giants. The light scattered off dust particles can be observed using non-coronagraphic AO polarimetry, requireing high spatial resolution due to small stellar diameters ($\sim$50 mas for stars like $\alpha$~Ori, Mira, W~Hya, R~Dor, $\alpha$~Sco). This favors observations at small $\lambda$ with SPHERE-ZIMPOL, achieving resolutions of $\sim$20 mas. Alternative approaches include differential speckle polarimetry without AO \citep{Safonov2020}, and sparse aperture masking interferometry with SCExAO-VAMPIRES on Subaru, promising resolutions up to 10 mas \citep{Norris2015}.

The starlight that is polarized due to reflection on exoplanets is polarized, allows for the detection of exoplanets through high-resolution polarimetry. SPHERE-ZIMPOL was specifically optimized to search for exoplanets  around nearby bright stars \citep{Schmid2005} ($\alpha$~Cen A/B, Sirius, $\epsilon$~Eri, Altair, $\tau$~Cet) have been conducted \citep{Hunziker2020}. The challenge is extreme: a Jupiter-sized planet at 1 AU yields a polarimetric contrast of only $C_\mathrm{pol} = p_\mathrm{planet} \cdot I_\mathrm{planet}/I_\mathrm{star} \approx 2 \times 10^{-8}$ \citep{Buenzli2009}. For $\alpha$~Cen A, this could be reached with $t_\mathrm{exp} = 3.4$ h, though no detection was achieved.

\section{Conclusion \& Outlook}
\label{sec:Outlook}

Polarimetry has emerged as an essential method in planetary sciences and astronomy, providing insights into surface composition, atmospheric properties, and biosignatures beyond standard photometry. As highlighted in this chapter, within the framework of the Swiss National Centre of Competence in Research (NCCR) PlanetS, the utilization of polarimetry has facilitated research extending from detailed characterization of planetary surfaces and regolith simulants to the remote detection and characterization of Earth biosignatures.

The systematic studies of polarization at low phase angles have clarified regolith composition and particle interactions. Our laboratory experiments have shown how mineral mixtures significantly alter polarization phase curves compared to pure minerals. These results can potentially greatly impact asteroid classification and planetary surface analysis \citep{Spadaccia2022}. Additional experiments in combination with modeling and observations havee the potential of strongly aiding in interpreting the surface properties of regolith. Polarization measurements at small phase angles have also proven to be sensitive to frost formation, with distinct wavelength-dependent polarization variations observed during initial frost deposition stages, providing insights into frost thickness and crystalline structure \citep{Spadaccia2023}.

The polarization results obtained within the NCCR PlanetS on regoliths will also provide valuable input on new upcoming missions. Asteroid 99942 Apophis will fly by the Earth in April 2029, at a distance of 32’000 km only. Interactions with the Earth's gravity field will lead to significant changes in the spin state of the asteroid and local surface slopes, possibly resulting in redistribution of surface regolith and refreshment of the surface through loss of the space-weathered surface layer. The European Space Agency develops the RAMSES mission to rendezvous with Apophis two months before the encounter and accompany it as it flies by the Earth. The Colour and High-resolution Apophis Narrow-angle CamEra System (CHANCES) was selected by ESA for the mission and is currently under development at the University of Bern. The classification of asteroids by means of their measured light polarization properties and supporting laboratory characterizations of analogues led to the decision to include polarizers on the instrument. A 13-positions filter wheel provides multispectral and refocusing capabilities to the imager. Two of the wheel slots are foreseen to accommodate two polarizers in an orthogonal configuration to characterize the linear polarization of the surface at sub-meter resolution. This will be a unique opportunity to study in-situ and at high resolution how the properties of the asteroid surface affect the polarization of the reflected solar light. New laboratory experiments building upon the ones performed in the NCCR PlanetS \citep{Spadaccia2022} will be performed to better interpret the observations of Apophis by CHANCES. This development is also seen as a pathfinder toward the implementation of polarization-sensing capabilities on future imagers and spectrometers by advancing our technical capabilities as well as calibration and scientific interpretation of the results. 

Beyond atmosphereless bodies, polarization capabilities will be very useful to image the surface of Mars from orbit. Remote-sensing of the Martian surface is complicated by the constant presence of suspended dust in the atmosphere. Polarimetric dehazing techniques used on Earth \citep{liang2015} could be applied to orbital imaging on Mars (see Pommerol et al., this collection) to better distinguish the contributions to the signal from the surface and from the atmosphere and refine both the radiometric calibration of images and the retrieval of aerosols content and properties. Beside dehazing, polarization information would also help to determine the presence and properties of ices and frosts at the surface \citep{Spadaccia2023}

FlyPol has demonstrated reliable remote detection of circularly polarized biosignatures in dynamic remote sensing settings \citep{Patty2021, Mulder2022}. The systematic investigation of the circular polarizance of leaves also indicates that the signals are very robust and relatively insensitive to phase angle variations \citep{Patty2022}. The signals observed, however, are very subtle and the reliable detection of biosignatures outside our Solar System is therefore unlikely within the near future. Potential Solar System applications could include stellar occultation spectropolarimetry of the plumes of Enceladus and Europe \citep{Sparks2021, Grone2024} or even Venus \citep{Sparks2021}, and ground-based telescopic observations of Earth shine polarization.

In astronomical applications, adaptive optics (AO) systems combined with polarimetry have transformed high-contrast imaging, enabling observations of circumstellar disks, debris systems, and exoplanets otherwise impossible. Instruments such as VLT-SPHERE/ZIMPOL and Gemini Planet Imager illustrate how sophisticated AO and polarimetric integration can significantly enhance observational capabilities \citep{Schmid2018, Macintosh2014}. Looking forward, the integration of next-generation telescopes equipped with enhanced AO and polarimetric capabilities will likely yield even more refined observational data.

\section*{Acknowledgements}
This work has been carried out within the framework of the National Centre of Competence in Research PlanetS supported by the Swiss National Science Foundation under grants 51NF40\textunderscore205606. 
CHLP acknowledges receiving support from the Swiss State Secretariat for Education, Research and Innovation (SERI) under contract number MB22.00046. SPHERE-ZIMPOL observations of Europa and Callisto shown in Figure 7  were obtained via ESO's programmes 0101.C-0322(A) and 096.C-0514(A) respectively.

\bibliography{Setup/Library.bib}

\end{document}